# Application of PROFINET IO in Neutron Scattering Instruments

H. Kleines, A. Ackens and F. Suxdorf

*Abstract*—**The control systems of all neutron scattering instruments implemented by the Jülich Centre for Neutron Science (JCNS) are based on Siemens PLC technology. Historically PROFIBUS has been used for the communication of PLCs with supervisory computers, decentral periphery systems and other PLCs. Today, PROFINET IO is the most commonly used industrial real time Ethernet system and naturally supported by Siemens PLC systems. As a consequence, all new neutron instruments of JCNS are based on PROFINET IO. For the interfacing to supervisory computers based on CompactPCI, a CompactPCI carrier board for PC/104-Plus mezzanines has been developed, allowing the transparent use of the Siemens PC/104-Plus PROFINET IO controller CP1604 in CompactPCI systems. Linux is used as the operating system for supervisory computers and the software development employs the PROFINET IO-Base-API, commonly supported by Siemens PROFINET IO controllers. On top of this API, an application protocol for the communication with PLC-based motion systems has been implemented.**

*Index Terms*—**Compact PCI, Control System, PC104+, PROFINET IO**

## I. INTRODUCTION

JCNS, the neutron science division of Forschungszentrum Jülich, developed and operates 15 neutron instruments at its outstations ILL in Grenoble, the Spallation Neutron Source in Oak Ridge and at the FRM-II at Garching near Munich. Further neutron instruments are being developed for the future European Spallation Source in Lund. The control and data acquisition systems of these neutron instruments are responsible for a variety of tasks including detector readout, vacuum systems, cryogenic systems, sample environment devices and motion of many mechanical axes.

In order to reduce the overall development effort und the number of spare parts, these control systems are highly standardized and follow a common framework, the so-called Jülich-Münich standard, which is introduced in the next section. A guiding principle for the framework was to minimize the development efforts and to acquire as much from the market as possible. A key component of the framework is the consistent use of industrial technologies like PLCs,

fieldbus systems or decentral periphery in the front end. Main motivations are:

- low prices induced by mass market,
- inherent robustness,
- long term availability and support from manufacturer,
- powerful development tools.

Because of its strong market position, Siemens is the PLC vendor selected for all JCNS instruments. In the past the communication of PLCs with peripheral devices and with supervisory computers relied mainly on PROFIBUS, since it was the field bus standard naturally supported by Siemens PLCs.

In the recent years Ethernet-based industrial communication systems like PROFINET IO [?], Ethernet/IP, Ethercat, Powerlink or Modbus TCP became increasingly important for the communication with industrial devices and today PROFINET IO (see section III) is the most commonly used industrial Ethernet system and inherently supported by all newer PLCs from Siemens. Because of this JCNS did some effort to implement PROFINET IO in the control systems of its neutron instruments as described in sections IV and V.

## II. OVERVIEW OF THE JUELICH-MUNICH STANDARD

A control system according to the Jülich-Munich Standard is organized hierarchically into the following levels:

**Field level:** The field level is the lowest level, at which devices that are not freely programmable reside, like motor controllers, SSI controllers, PID controllers, analogue and digital I/O modules, or measurement equipment. For all industrial type of I/O modules PROFIBUS DP based decentral periphery is normally used. Siemens ET200S is the preferred one. The ET200S modules 1STEP and 1STEP-DRIVE are the predominantly used stepper motor controllers.

**Control level:** The control level resides on top of the process level. Devices at the control level are freely programmable. They must meet real time requirements and guarantee robust operation in a harsh environment. At the control level Siemens S7 PLCs are used, because they dominate the European market.

**Process communication:** Process communication covers the communication of devices at the field and control level with supervisory controllers or computers. For lab equipment GPIB and proprietary RS232/RS485 connections are unavoidable. For industrial automation, normally PROFIBUS DP is chosen. It is the dominating fieldbus systems in Europe and naturally supported by S7 PLCs and many other devices. A major reason for their success is the technological and functional scalability based on a common core as well as the programming model, which easily maps to PLC operation.

**Experiment Computer:** For economic reasons, all experiment computers should be PCs. Linux, being well established in the scientific community, is the only supported operating system. CentOS is the preferred distribution at JCNS. Direct device access is typically implemented on industrial PCs, mainly CompactPCI systems. CompactPCI allows deploying a variety of existing software in a mechanically more robust platform that fits into 19" racks.

**Middleware:** Since the framework aims at an inherently distributed system, software support for the transparent distribution of services between systems is required. For this purpose TACO [2] originally has been selected as the middleware system. TACO is a client-server framework developed for beam line control at the ESRF in Grenoble. In a TACO environment each device or hardware module is controlled by a TACO server. Meanwhile TACO has been replaced by its object—oriented successor TANGO [3] at most JCNS instruments.

**Application level:** On the client side originally a high degree of freedom was allowed ranging from python based pure scripting applications to more static GUI-applications in C++. At most JCNS instruments these applications have been replaced by the standardized measurement program NICOS [4] during the last few years. NICOS offers scripting in python and in a simpler command language as well as a configurable GUI for graphical user operation. Functionalities comprise electronic logbook, history plots, detector data plots,…..

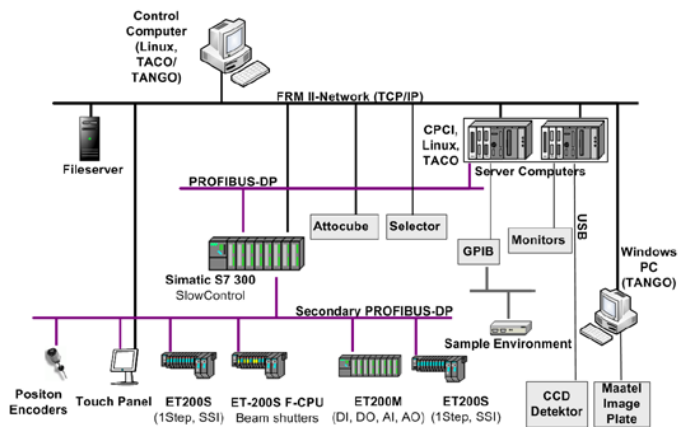

Fig. 1 Physical architecture of the BIODIFF control and DAQ system

As a typical example, Fig. 1 shows the physical architecture of the control and DAQ system of the neutron diffractometer BIODIFF[5]. It is implemented as a distributed system with a hierarchical architecture. On top, the so-called control computer resides with all application software. Via the experiment network the control computer accesses the "server computers", to which all front end systems (detectors, position encoders, motor controllers, digital IOs, analogue IOs, …) are attached. On the "server computers" TANGO servers are running, which access the peripheral devices via dedicated device drivers.

The "slow control" peripherals are indirectly connected to the "server computers" via a PROFIBUS segment with the main S7-300 PLC equipped with the failsafe CPU 319F-3 PN/DP. This CPU contains both the "standard" program as well as the failsafe program for the personal protection in parallel.

Stepper motor controllers and SSI modules as well as digital and analogue I/Os reside in three modular ET200 decentral periphery systems, which are connected to the PLC via an additional subordinate PROFIBUS segment.

### III.  PROFINET IO

Originally defined by PI (PROFIBUS and PROFINET International), PROFINET IO is now internationally standardized in IEC 61158 and IEC 61785. It is based on 100 Mbit/s Ethernet supporting wireless, copper and optical media in full conformance to the corresponding IEEE standards. As a consequence, a PROFINET communication interface can be used simultaneously for PROFINET real time communication as well as for standard internet communication – e.g. for web services. PROFINET IO defines several conformance classes in order to support soft real time as well as hard real time communication, called IRT (isochronous real time) mode. In IRT mode, communication cycle times down to 31.25 µs with a jitter of less than 1 µs are supported. IRT mode relies on PTP (Precision Time Protocol) according to IEEE 1588 for time synchronization.

In order to facilitate the easy migration from PROFIBUS to PROFINET, it shares basic architectural principles with PROFIBUS DP. Similar to PROFIBUS DP, PROFINET IO aims at application scenarios, where a central station communicates with decentral field devices. As a consequence, each station in a PROFIBUS IO system can take one of the following roles according to Fig. 2:

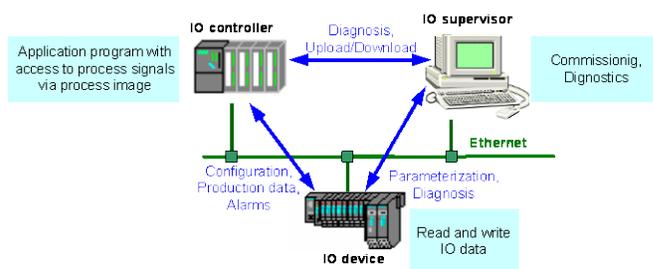

Fig. 2. PROFINET IO communication model

- IO controller: The IO controller represents an intelligent central station, like a PLC. It is responsible for the configuration and parameterization of its associated devices and controls the transfer of process data.
- IO device: The IO device represents a field device, like an analogue input unit. It cyclically transmits collected process data to the IO controller and vice versa. It also provides diagnostic or alarm information to the IO controller.

- IO supervisor: The IO supervisor represents an engineering station for programming, configuration or diagnostics, e.g. a PLC programming tool.

Similar to PROFIBUS DP, PROFINET IO is based on a consistent model of the IO device structure and capabilities, described by so-called GSDML files. An IO device may be modular and is composed of one or more so-called slots, which may have subslots. Each slot or subslot represents an IO module and has a fixed number of input and output bits. The input data of the IO device is the sequence of all inputs of slots and subslots, according to their position in the device. The same holds for the output data. Also all diagnostic or alarm data reference slots or subslots.

Similar to PROFIBUS DP, the transfer of process data relies on a consumer/producer model with cyclic data exchange between an IO controller and its IO devices. As illustrated in Fig. 3 the transmission occurs from and to prepared buffer areas. The buffer areas are read and written by the application processes in IO controller and in IO device synchronously or asynchronously to the communication cycle. This memory-based communication model corresponds in a very natural way to internal PLC operation and makes software implementation of data exchange extremely simple.

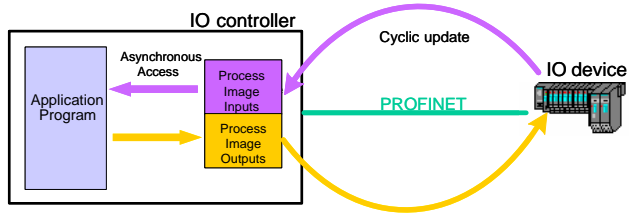

Fig. 3. Cyclic message exchange

## IV. IMPLEMENTATION OF PROFINET IO FOR THE COMMUNICATION BETWEEN PLCS AND FIELD DEVICES

As a first step PROFINET IO was introduced for the communication between PLCs and ET200 field devices. This was basically straight forward, since PROFINET IO is well supported by the S7-300 and S7-1500 PLC families as well as by the decentral periphery systems ET200S, ET200pro, ET200M, ET200SP and ET200MP used by JCNS. Tests in the lab showed that this was simply sufficient to change the ET200 interface modules from PROFIBUS to PROFINET (e.g. from IM151-1 to IM151-3 PN in the case of the ET200S). In the case of the PLC the internal PROFINET interface of the CPU (e.g. in the case of the CPUs 315-2 PN/DP or 319F-3 PN/DP) could be used instead of the internal PROFIBUS interface. Communication modules like the CP343-1 were used only for the communication with the supervisory computers. Due to the conceptual similarity between PROFINET and PROFIBUS process data transfer – both basically do read and write access to the process IO memory image - only minor syntactical changes in the PLC software had to be made, which are well described in a PROFIBUS-to-PROFINET migration guide form Siemens. Of course, the hardware configuration in the development environment TIA portal is different, e.g. regarding address format or communication parameters. Based on the positive results in the lab, extensions and modifications of existing instruments are usually done with PROFINET IO, now. This proved that the implementation of mixed PROFINET/PROFIBUS installations is straight forward, which allows very flexible architectures. Also a flexible mix of the PLC families S7-1500 and S7-300 and the different ET200 families were was easily possible - only in some cases a firmware update of modules was required. PROFIENT IO is well-supported by different PLC vendors as well as by device vendors (e.g. for servo drives or position encoders), too.

For the implementation of new control systems, now usually pure PROFINET systems are implemented. As an example, Fig. 4 shows the physical architecture of the motion subsystem that has been implemented for the neutron radiography instrument ANTARES[6].

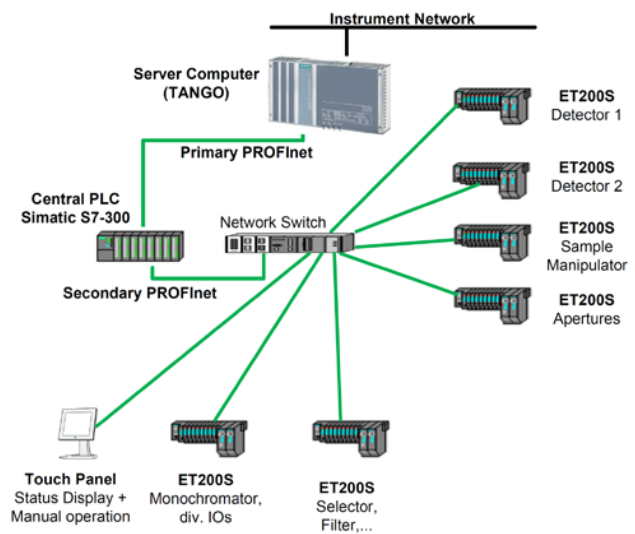

Fig. 4 Physical architecture of the ANTARES motion subsystem

## V. IMPLEMENTATION OF PROFINET IO FOR SUPERVISORY COMPUTERS

For the implementation of the PROFINET IO interface in supervisory computers a variety of controllers for PCI, CompactPCI and PCIe with Linux support are available from different vendors. Additionally, there are industrial computers on the market with PROFINET IO directly integrated on the main board – as in the Siemens box PC shown in the ANTARES motion subsystem shown in Fig. 4 – as well as pure software implementations for standard Ethernet network interface cards, e. g. the PROFINET driver in source code, which also supports Linux.

For CompactPCI, which is required for the JCNS instruments, several well established vendors – like Hilscher or Kunbus – offer PROFINET IO controller implementations. Since PROFINET IO requires a consistent configuration of the controller (in our case the PC) and the device (in our case the PLC) we would have to use tools from different vendors for this purpose, which is quite error-prone. So we intended to use a Siemens product also for the controller, which could be configured directly from the TIA Portal PLC project. Since no CompactPCI controller is available from Siemens, we selected the product CP1604 in PC104+ formfactor. The CP1604 supports PROFINET IO controller as well as device functionality and comes with Linux support, including device driver, utilities and examples programs in source code.

For the integration of the CP1604 we implemented a generic CompactPCI base board for PC104+ mezzanines, as shown in Fig. 5.

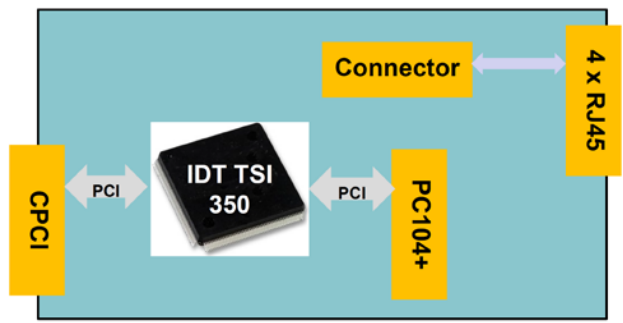

Fig. 5 Simplified block diagram of the new CompactPCI carrier for PC104+

Core of the board is the IDT TSI 350 PCI bridge chip, supporting 32 Bit / 66 MHz PCI. As shown in Fig. 6, the 4 port RJ45 connector required for CP1604 has been directly integrated on the base board and is interfaced via a connector to the mezzanine, thus avoiding a ribbon cable. Due to the PC104+ connectors, two CompactPCI slots are required for a standard mezzanine. For a single slot configurations, we exchanged the female/male connectors on the CP1604 by a single female connector.

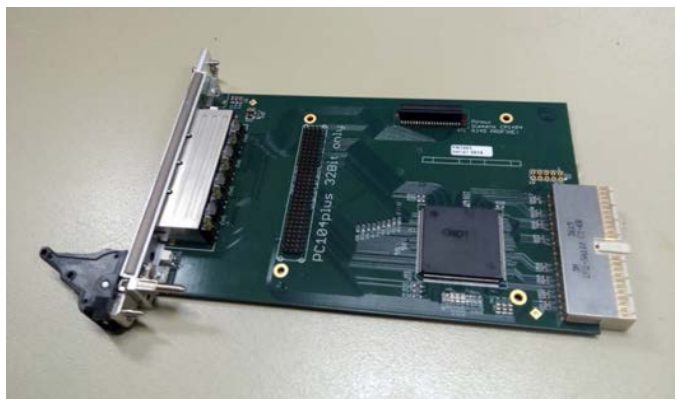

Fig. 6 Photo of the new CompactPCI carrier for PC104+

## VI. Conclusions and Outlook

PROFNET IO has been successfully implemented in JCNS neutron instruments as a successor of PROFIBUS DP. Introduction of PROFINET IO to the devices at the field level and the PLCs at the control level was straight forward with moderate effort due to the good product support by the vendor Siemens and the conceptual similarity to PROFIBUS DP. The possibility to almost arbitrarily mix PROFIBUS and PROFINET as well as different PLC and ET200 families is considered as a major advantage allowing very flexible architectures and an easy migration to newer products and technologies.

For the implementation of the PROFINET IO interface for CPCI-based supervisory computers a CPCI carrier module for PC104+ was implemented that allowed the transparent integration of the PC104+ mezzanine CP1604 from Siemens as an IO controller. Future work will evaluate the PROFINET Driver as am example for a pure software implementation of PROFINET IO. for standard Ethernet controllers, also available for Linux. Comparative performance measurements are planned, too.